\documentclass[journal,doublecolumn,10pt]{IEEEtran}
\usepackage{graphicx,cite,epsfig,amssymb,amsmath,color}
\usepackage{amssymb}
\usepackage{amsmath}
\usepackage{graphicx}
\usepackage{cite}
\usepackage{enumerate}
\usepackage{epstopdf}
\usepackage{chngpage}
\usepackage{array}
\usepackage{footnote}
\usepackage{subfigure}
\usepackage{stfloats}
\usepackage{bm}
\usepackage{color}
\usepackage{setspace}
\makesavenoteenv{tabular}
\begin{document}

	
	
	\title{Energy-Efficient Joint Offloading and Wireless Resource Allocation Strategy in Multi-MEC Server Systems}
	
	\author{\IEEEauthorblockN{Kang Cheng\IEEEauthorrefmark{1}, Yinglei Teng\IEEEauthorrefmark{1}, Weiqi Sun\IEEEauthorrefmark{1}, An Liu\IEEEauthorrefmark{2}, \IEEEmembership{Senior Member, IEEE}, Xianbin Wang\IEEEauthorrefmark{3}, \IEEEmembership{Fellow, IEEE}\\}

\IEEEauthorblockA{\IEEEauthorrefmark{1} Beijing Key Laboratory of Space-ground Interconnection and Convergence, Department of EE, BUPT\\ \normalsize{Email: chengkang@bupt.edu.cn, lilytengtt@gmail.com, weiqisun@bupt.edu.cn}\\}
\IEEEauthorblockA{\IEEEauthorrefmark{2} Department of ECE, HKUST, HK,  \normalsize{Email: eewendaol@ust.hk }}
\IEEEauthorblockA{\IEEEauthorrefmark{3} Department of ECE, UWO, CA,  \normalsize{Email: wang@eng.uwo.ca}}

\thanks{This work was supported by National Natural Science Foundation of China under Grant No. 61771072, and Beijing Natural Science Foundation under Grant No.L171011.
}}
\vspace{-5mm}	

	\maketitle
	\begin{abstract}
Mobile edge computing (MEC) is an emerging paradigm that mobile devices can offload the computation-intensive or latency-critical tasks to the nearby MEC servers, so as to save energy and extend battery life. Unlike the cloud server, MEC server is a small-scale data center deployed at a wireless access point, thus it is highly sensitive to both radio and computing resource. In this paper, we consider an Orthogonal Frequency-Division Multiplexing Access (OFDMA) based multi-user and multi-MEC-server system, where the task offloading strategies and wireless resources allocation are jointly investigated. Aiming at minimizing the total energy consumption, we propose the joint offloading and resource allocation strategy for latency-critical applications. Through the bi-level optimization approach, the original NP-hard problem is decoupled into the lower-level problem seeking for the allocation of power and subcarrier and the upper-level task offloading problem. Simulation results show that the proposed algorithm achieves excellent performance in energy saving and successful offloading probability (SOP) in comparison with conventional schemes.
	
	\vspace{0.2cm}
	\end{abstract}	
	\IEEEpeerreviewmaketitle
    \vspace{0.5cm}
	\section{Introduction}
\begin{spacing}{0.95}
With the advent of 5G era, the explosive growth of smart devices especially intelligent mobile phones and Internet of Things (IoT) devices promotes user's requirement for the (ultra) low-delay high-quality services, e.g., mobile gaming and augmented reality/virtual reality (AR/VR). Computing demand is prominent than ever that it frequently exceeds what local mobile devices can deliver \cite{Refer1}. Cloud computing is believed to have powerful computing abilities, however, it is largely trapped by the limited backhaul. To alleviate the backhaul pressure and provide the lower delay, mobile edge computing (MEC) is presented as a new paradigm that attracts attentions from academia to industry. Its main feature is to harvest the vast amount of the idle computation power and storage space distributed at the network edges to perform computation-intensive and latency-critical tasks at mobile devices \cite{Refer2}.

\indent Both radio and computing resources are particularly important for the performance of task offloading: the former determines the data rate and energy consumption in transmission process while the latter restricts the computing time and energy consumption of tasks offloaded to an MEC server \cite{Refer3}. Conventional studies on wireless resource allocation focus on the spectrum and energy efficiency, while computational resource optimization addresses much on the {distributed computing \cite{Refer4} and code partition \cite{Refer6}}. However, in the both types of resource-constrained MEC system, the transmission and computation processes are coupled especially when the system level delay or energy metric is targeted. A tradeoff should exist between the  \textit{multi-user diversity} providing the gain through the channel fading with the multi-user selection and  \textit{multi-MEC diversity} exploring the computation opportunity with multi-MEC selection. However, the combined optimization for MEC systems becomes more complicated with the increasing amount of users and MEC servers. Therefore, it is essential to develop effective offloading strategies for the MEC systems characterized by multiple users and servers.

\indent Recent researches on MEC mainly focus on the offloading strategy or the joint radio and computing resources optimization problem. However, a large body of existing works just consider the MEC systems with either single user or single MEC server.
In \cite{Refer3}, the authors consider a cloudlet in an Orthogonal Frequency-Division Multiplexing Access (OFDMA) system with multiple mobile devices and propose a joint scheduling algorithm that allocates both radio and computing resources coordinately. \cite{Refer8} optimizes the radio and computing resources in a MIMO multicell system where multiple mobile users ask for computation offloading to a common cloud server. An integrated framework for computation offloading and interference management is proposed in \cite{Refer9} for the single MEC server system. In \cite{Refer10}, a novel user-centric mobility management scheme is developed to maximize the edge computation performance for the single user while keeping the user’s communication energy consumption below a constraint in the multi-MEC-server environment. As far as we know, this is the first work that exploits the joint optimization of task offloading strategy and radio resource allocation problem in the multi-users multi-MEC-server systems.

\indent In this paper, we propose an energy-efficient joint offloading and wireless resource allocation strategy (EEJS) that jointly optimizes the offloading strategy and radio resources to reduce the total energy consumption of the whole MEC system. Due to the non-convex and NP-hard feature for the joint optimization problem, the original problem is converted into a bi-level optimization problem. Given the offloading strategy, the lower-level problem turns to minimizing the transmission energy for the computing task offloading. To be specific, we approximate the lower-level problem with the latency restriction and find the optimal power and subcarrier allocation with Lagrangian method. In the upper-level problem, the optimal offloading strategy with respect to minimal total energy consumption objective is selected from all offloading strategies according to the condition and constraints. Comparing with conventional offloading strategies, simulation results reveal that the proposed EEJS advances in both system energy consumption and successful offloading probability (SOP), which indicates it achieves the good tradeoff between the multi-user diversity and multi-MEC diversity.

	\vspace{-8mm}
	\end{spacing}	
	\section{System  Model}
	\begin{spacing}{0.95}
	\begin{figure}[!t]\label{fig1}
		\centering
		\includegraphics[width=7.0cm,height=4.5cm]{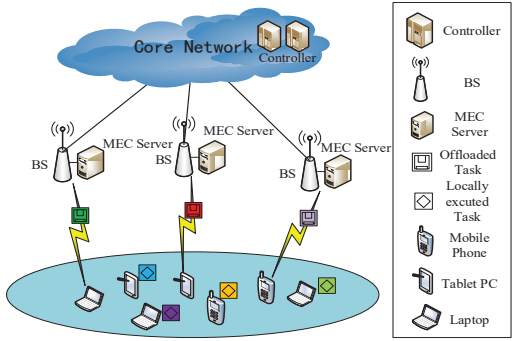}
\vspace{-0.3cm}
		\caption{A mobile edge computing system with multiple mobile devices and multiple-MEC servers.}
	\end{figure}
\vspace{-0.1em}
In this section, we analyze the required energy for task offloading and computation, and formulate the system energy minimization problem.


  As shown in Fig. 1, we consider a snapshot of MEC system with multiples users and multiple available MEC servers, while OFDMA is utilized as the uplink transmission mechanism. Multiple MEC servers are equipped with certain computation capability and deployed at the base station (BS) to provide task execution. We assume the MEC servers work independently and connects with the core network via the fiber. 	
Assume the central processing units (CPUs) of $K$ MEC servers are idle at the current time, whose set is indicated by ${\cal K} = \left\{ {1,2,...,K} \right\}$. There are $N$ available subcarriers for uplink wireless transmission and ${\cal N} = \{ 1,2,...,N\}$. The bandwidth of each subcarrier is $B_N$. Let ${\cal I} = \left\{ {1,2,...,I} \right\}$ denote the set of active users, each having one task, which means that there are $I$ pending tasks. Without ambiguity, we also use $i$ as the task indicator, and each task is described by a three-field notation $A({D_i},{\tau _i},{X_i})$. This commonly-used notation contains the information of the task input-data size ${D_i}$ (in bits), the requested completion deadline ${\tau _i}$ (in second) and the computation workload/intensity ${X_i}$ (in CPU cycles per bit). Note that these parameters are related to the nature of the applications and can be estimated through task profilers \cite{Refer11},\cite{Refer12}. Besides, we assume all the MEC servers have enough capacity for task execution and will execute the tasks till finished once assigned.

For user $i$, the frequency of local CPU is ${f_{i,loc}}$, and the maximal transmission power of all users is set same and denoted by ${P^m}$. The CPU frequency of MEC server $k$ is denoted by ${f_{k,ser}}$. Let ${\bf{W}} = \{ {w_{i,n,k}}|{w_{i,n,k}} \in \{ 0,1\} ,i \in {\cal I},n \in {\cal N},k \in {\cal K}\}$ denote the subcarrier allocation matrix, where  ${w_{i,n,k}} = 1$ means the subcarrier $n$ is allocated to the user $i$ whose task is offloaded to the MEC server $k$ while ${w_{i,n,k}} = 0$ means not. The subcarrier power allocation matrix is denoted by ${\bf{P}} = \{ {p_{i,n,k}}|{p_{i,n,k}} \in [0,{P^m}],i \in {\cal I},n \in {\cal N},k \in {\cal K}\}$, where ${p_{i,n,k}}$ is the power of user $i$ whose task is offloaded to the MEC server $k$ allocated on subcarrier $n$. Let ${\bf{G}} = \{ {g_{i,n,k}},i \in {\cal I},n \in {\cal N},k \in {\cal K}\}$ denote the channel-gain matrix of subcarrier, and each element ${g_{i,n,k}}$ denotes the channel-gain between the remote server $k$ to user $i$ on the subcarrier $n$. Meanwhile, we assume a flat fading environment such that the channel-gain matrix ${\bf{G}}$ remains constant during once scheduling process. Moreover, the system noise accords to the zero expectation Gaussian distribution whose variance is denoted by ${\delta} ^2$.

Considering the total required CPU cycles for completing the task $i$ given by ${X_i}{D_i}$, the calculation time cost is defined as the required CPU cycles divided by the CPU frequency.
Since the task is to be executed locally or remotely, the time cost is discussed as follows.
	
	
The time cost for local execution is determined by local computing capability, thus the local execution time for user $i$ is given as
\vspace{-3mm}
\begin{small}
	\begin{equation}\label{TLocalE}
	{T_{i}^l} = \frac{{{D_i}{X_i}}}{{{f_{i,loc}}}}.
	\end{equation}
\end{small}
\setlength{\abovedisplayskip}{2pt}
   For the offloading task, the execution time consists of two parts, i.e., the transmission time spent on the wireless channel and the actual computing time cost by the remote MEC servers. To be specific,

   i) \emph{Transmission}:  In view of the OFDMA mechanism, interference is ignored in virtue of the
exclusive subcarrier allocation. For user ${i}$ who offloads task to the MEC server ${k}$, the aggregated data rate is expressed as
   \begin{small}
   \begin{equation}\label{DataRate}
   {R_{i,k}} = {B_N}\sum\limits_{n \in \cal{N}} {{w_{i,n,k}}{{\log }_2}(1 + \frac{{{g_{i,n,{\rm{ }}k}}{p_{i,n,k}}}}{{{\delta ^2}}}){\rm{    }}}.
   \end{equation}
   \end{small}
   Then, the time cost for transmitting task ${i}$ to MEC server ${k}$ is given by
   \vspace{-3mm}
   \begin{small}
   \begin{equation}\label{Tremotran}
   T_{i,k}^t = \frac{{{D_i}}}{{{R_{i,k}}}}.
   \end{equation}
\end{small}
   ii) \emph{Computing}: We assume the non-preemptive CPU allocation, which assigns a time slot to one user each time until its task is completed. Since only idle MEC servers are considered for once scheduling, the waiting time for queue is not involved. Then, the time cost for computing task ${i}$ in MEC server ${k}$ depends on the calculation intensity and CPU frequency, which is given by

  \vspace{-4mm}
   \begin{small}
   \begin{equation}\label{Tremocomp}
   T_{i,k}^c = \frac{{{D_i}{X_i}}}{{{f_{k,ser}}}}.
   \end{equation}
   \end{small}
  Accordingly, the total time cost for the remote execution is given by
  \vspace{-2mm}
  \begin{small}
  \begin{equation}\label{Tremototal}
  T_{i,k}^r = T_{i,k}^t + T_{i,k}^c = \frac{{{D_i}}}{{{R_{i,k}}}} + \frac{{{D_i}{X_i}}}{{{f_{k,ser}}}}.
  \end{equation}
\end{small}
    Also, the energy consumption is presented separately with respect to the different execution modes as below.

    i) \emph{Local execution}: According to \cite{Refer13}, given the running frequency ${f_{i,loc}}$, the energy consumption on user $i$ during each CPU cycle is ${k_0}f_{i,loc}^2$, where $k_0$ is a constant related to CPU of mobile user. Thus, the energy consumption of task ${i}$ for local execution is given by
    \begin{small}
    \begin{equation}\label{Elocal}
    E_i^l = {k_0}f_{i,loc}^2{D_i}{X_i}.
    \end{equation}
\end{small}
   ii) \emph{Remote execution}: In the case of offloading, similarly, it incorporates the transmission energy used to send the input data $D_i$ to the helping MEC server $k$ and the energy consumption on the MEC server $k$ for computing. Here, the transmission energy for passing back the computation results is ignored as in \cite{Refer13, Refer14, Refer15}, since the amount of output data is usually much less than that of the input data.

   The energy consumption for transmitting task ${i}$ to MEC server ${k}$ is given by
   \vspace{-1.5mm}
   \begin{small}
   \begin{equation}\label{Eremotran}
   E_{i,k}^t = \sum\limits_{n \in \cal {N}} {{w_{i,n,k}}{p_{i,n,k}}} \frac{{{D_i}}}{{{R_{i,k}}}}.
   \end{equation}
   \end{small}
   \setlength{\abovedisplayskip}{3pt}
   Similar with the local execution, the energy consumption for computing task ${i}$ in MEC server ${k}$ can be denoted by
   \begin{equation}\label{Eremocomp}
   E_{i,k}^c = {k_1}f_{k,ser}^2{D_i}{X_i},
   \end{equation}
   where $k_1$ is also constant and related to the CPU of MEC server ${k}$.
   Then, the total energy consumption for remote execution is given by
   \vspace{-2mm}
   \begin{small}
   \begin{equation}\label{Eremosum}
E_{i,k}^r \!= \!E_{i,k}^c \!+\! E_{i,k}^t\! =\! {k_1}f_{k,ser}^2{D_i}{X_i}\! +\! \sum\limits_{n \in \cal {N}} {{w_{i,n,k}}{p_{i,n,k}}} \frac{{{D_i}}}{{{R_{i,k}}}}.
   \end{equation}
\end{small}
\setlength{\abovedisplayskip}{3pt}
{Rather than working on the offloading strategy of ``to offload or not" which has been intensively addressed in \cite{Refer3}, \cite{Refer9}, in this paper, focusing on the multi-MEC multi-user cases, we only devise a simplified ``to offload or not" offloading decision, but propose the wireless transmission aware joint offloading strategy solving the problem of ``offload to which one" (offloading allocation).}
Here, the simplified offloading strategy which decides if the task is executed locally or remotely is given as follows,
   \begin{equation*}\label{funpd}
   \left\{ {\begin{array}{*{1}{l}}
{{b_{i,0}}{\rm{ = }}1\hspace{0.6mm}:\left\{ {E_i^l < {E_0}} \right\} \cap \left\{ {T_i^l < {\tau _i}} \right\}}\\
{{b_{i,0}}{\rm{ = }}0:\hspace{0.6mm}\text{otherwise}}
\end{array}} \right.\hspace{-3.3mm},\forall i \in {\cal I}.
  \end{equation*}
\setlength{\abovedisplayskip}{3pt}
Here, ${b_{i,0}}{\rm{ = }}1$ means task ${i}$ is executed locally while ${b_{i,0}}{\rm{ = }}0$ otherwise. Meanwhile, let ${b_{i,k}}$ be the indicator of whether task ${i}$ is offloaded to MEC server ${k}$, and ${{b_{i,k}}{\rm{ = 1}}}$ indicates task ${i}$ is offloaded to MEC server ${k}$ while ${{b_{i,k}}{\rm{ = 0}}}$ means not. To avoid excessive energy consumption at terminals, we set an energy threshold ${E_0}$ restricting the maximal local consumption. Meanwhile, ${\tau _i}$ is the delay threshold for task $i$.
Hence, only the above two conditions are satisfied simultaneously, a computing task can be treated as the local profitable services and executed locally, otherwise, we decide the multi-MEC offloading strategies utilizing the proposed EEJS in the next section.
\vspace{-3mm}

   \section{Energy Consumption Minimization Problem}
  Consider the task offloading users whose set is updated as  ${{\cal I}'}, {{\cal I}'} \subset {{\cal I}}$. Next, we will deal with the EEJS deciding to which MEC server each pending task is to be offloaded as well as solving the efficient wireless resource allocation. Concerning with the system energy consumption, the joint optimization of the offloading and wireless resource allocation problem is formulated as,

  \vspace{-2mm}
\begin{small}
\begin{equation*}\label{funcP}
\begin{aligned}
{\cal P}&:\mathop {min}\limits_{{\bf{b,W,P}}} {\rm{  }}F({\bf{b}},{\bf{W}},{\bf{P}}) = \sum\limits_{i\in \cal{I'}} {\sum\limits_{k \in \cal{K}} {{b_{i,k}}E_{i,k}^r} } \\
s.t.& C1:{b_{i,k}} \in \{ 0,1\} ,\forall i \in {{\cal I}'},{\rm{ }}\forall k \in {\cal K}\\
& C2:\sum\limits_{i\in \cal{I'}} {{b_{i,k}}}  \le 1,{\rm{ }}\forall k \in {\cal K},\quad C3:\sum\limits_{k \in \cal{K}} {{b_{i,k}}}  = 1,{\rm{ }}\forall i \in {{\cal I}'}\\
& C4:\sum\limits_{k \in \cal{K} } {\sum\limits_{n \in \cal{N}}{{w_{i,n,k}}} } {p_{i,n,k}} \le {P^m},\forall i \in {\cal I}'\\
& C5:{w_{i,n,k}} \in \left\{ {0,1} \right\},{\rm{ }}\forall i \in {{\cal I}'},\forall n \in {\cal N},\forall k \in {\cal K}\\
& C6:\sum\limits_{i\in \cal{I'}} \sum\limits_{k \in \cal{K}} {{w_{i,n,k}}}  = 1,{\rm{ }}\forall n \in {\cal N}\\
& C7:{T_i} = \sum\limits_{k \in \cal{K}} {{b_{i,k}}\left( {\frac{{{D_i}}}{{{R_{i,k}}}} + \frac{{{D_i}{X_i}}}{{{f_{k,ser}}}}} \right)}  \le {\tau _i},{\rm{ }}\forall i \in {{\cal I}'}\\
\end{aligned}
\end{equation*}
\end{small}where ${{\bf{b}}=\{b_{i,k},\forall i\in{\cal{I'}},\forall {k}\in\cal{K}\}}$, ${\bf{W}}=\{w_{i,n,k},\forall i\in{\cal{I'}},\forall{n}\in{\cal{N}},\forall {k}\in\cal{K}\}$, ${\bf{P}}=\{p_{i,n,k},\forall i\in{\cal{I'}},\forall{n}\in{\cal{N}},\forall {k}\in\cal{K}\}$.
Constraint $C1$ shows that ${b_{i,k}}$ is the binary variable indicating whether task ${i}$ is offloaded to MEC server ${k}$. $C2$ ensures that each MEC server can only accepts no more than one task at a time. $C3$ means each task is offloaded to only one MEC server exclusively. $C4$ is the maximal power budget for each user. In addition, constraint $C5-C6$ are the subcarrier allocation constraints ensuring that each subcarrier is assigned exclusively to one user and $C7$ enforces the corresponding hard deadline on each offloaded task.

This resource allocation problem is a mixed-integer nonlinear programming (MINLP) problem, which in general is NP hard.  
In the following, we will propose efficient algorithms to solve this problem with the optimal performance. Based on the idea of multi-objective hierarchical optimization, the problem ${\cal P}$ could be equivalently transformed into ${\cal P}1$ as follows:

  \vspace{-2mm}
\begin{small}
\begin{equation*}\label{funcP1}
\begin{aligned}
{\cal P}1&:\mathop {min}\limits_{{\bf{b,W,P}}} {\rm{  }}F({\bf{b}},{\bf{W}},{\bf{P}}) = \sum\limits_{i \in {\cal{I'}}} {\sum\limits_{k \in \cal {K}} {{b_{i,k}}E_{i,k}^r} } \\
s.t.& C1 - C7\\
& C8{\rm{: (}}{\bf{W,P}}{\rm{)}} \in {\rm{arg  }}\mathop {{\rm{min}}}\limits_{{\bf{b}}\in \cal{B}} {\rm{  }}F({{\bf{b}}},{\bf{W}},{\bf{P}})
\end{aligned}
\end{equation*}
\end{small}where ${{\cal{B}}}$ denotes the feasible region of $\bf{b}$. Note that ${\cal{P}}1$ involves two embedded problems.
On one hand, the energy consumption of the entire system can be calculated only when the offloading strategy is known. On the other hand, the offloading strategy ${\bf{b}}$ is inversely influenced by subcarrier power ${\bf{P}}$ and subcarrier allocation strategy ${\bf{W}}$. Because that the characteristics of mutual restriction meet the requirements of the bi-level optimization problem (BLP), we are motivated to tackle the optimization problem ${\cal P}$ through the bi-level optimization problem.
\vspace{-4mm}
\section{Energy-Efficient Joint Strategy Algorithm}
\vspace{-1mm}
\begin{small}
\begin{figure*}[ht]
\begin{align*}
\frac{{\partial \Re ({\bm{\alpha }},{\bm{\beta }},{\bf{W}},{\bf{P}^*(b)})}}{{\partial {w_{i,n,k}}}}\!\!\left| {_{{w_{i,n,k}}  \!= \! w_{i,n,k}^*}} \right. \!\!\!\! =  \!\left( {b_{i,k} {\chi _i}  \!+ \! {\beta _{i}}} \right)p_{i,n,k}^* \!-  \!{\alpha _{i,k}}{B_N}{\log _2}\!\left(\! {1 \! + \! \frac{{{g_{i,n,k}}p_{i,n,k}^*}}{{{\sigma ^2}}}} \!\right)
\!\!= \!{\phi _{i,n,k}}\left\{ \begin{array}{l}
 \!> \!0{\rm{   }}, \!\text{if}  w_{i,n,k}^*{\rm{  \!= \! 0 }}\\
 \!= \!0{\rm{   }}, \!\text{if}  w_{i,n,k}^* \in \left( {{\rm{0,1}}} \right){\rm{ }}\\
 \!< \!0{\rm{   }}, \!\text{if}  w_{i,n,k}^*{\rm{  \!= \! 1 }}
\end{array} \right.
\tag{14}
\end{align*}
\hrulefill 
\end{figure*}
\end{small}
In view of the problem ${\cal P}1$, we adopt bi-level optimization approach to solve the original problem ${\cal P}$. 
Firstly, given the task offloading strategy ${{\bf{b}}}$, the optimal power allocation ${\bf{P}}$ and subcarrier allocation strategy ${\bf{W}}$ are solved by function $F({{\bf{b}}},{\bf{W}},{\bf{P}})$.
Then, according to the optimal power allocation ${\bf{P}^*(b)}$ and subcarrier allocation strategy ${\bf{W}^*(b)}$ which have been found in lower-level problem, the optimal task offloading strategy ${\bf{b}^*}$ is solved by the function $F({\bf{b}}, {\bf{W}^*(b)},{\bf{P}^*(b)})$ in the upper-level problem.
\end{spacing}
\vspace{-4mm}
\subsection{Lower-level Problem}
Given the task offloading strategy ${{\bf{b}}}$, since the total computation energy consumption becomes the known quantity, the lower-level problem can be written in the form of ${\cal P}2$ as below.
\vspace{-1mm}
\begin{small}
\begin{equation*}\label{funcP2}
\begin{aligned}
{\cal P}2&:\mathop {min}\limits_{{\bf{W,P}}} {\rm{  }}F' ({\bf{b}}{},{\bf{W}},{\bf{P}}) = \sum\limits_{i \in {\cal{I'}}} {\sum\limits_{k \in \cal{K}} {b_{i,k}} } \sum\limits_{n \in \cal{N}} {{w_{i,n,k}}{p_{i,n,k}}} \frac{{{D_i}}}{{{R_{i,k}}}}\\
s.t.& C4 - C6,\\
& C9:\sum\limits_{k \in \cal{K}} {{b_{i,k}}\frac{{{D_i}}}{{{R_{i,k}}}}}  \le {\chi _i}{\rm{ }},\\
&{\chi _i} = {\tau _i} - \sum\limits_{k \in \cal{K}} {{b_{i,k}}\frac{{{D_i}{X_i}}}{{{f_{k,ser}}}}} {\rm{ , }}\forall i \in {{\cal I}'}{\rm{ }}.
\vspace{-2mm}
\end{aligned}
\end{equation*}
\end{small}
\setlength{\abovedisplayskip}{3pt}
Due to the discreteness of ${{\bf{b}}}$, constraint $C9$ can be equivalently transformed into multiple parallel constraints as $C10$,
\begin{small}
   \begin{equation*}\label{Convert1}
  C10:{b_{i,k}}\frac{{{D_i}}}{{{R_{i,k}}}} \le {\chi _i}{\rm{ }},{\rm{ }}\forall i \in {{\cal I}'}{\rm{,}}\forall k \in {\cal K}.
  \end{equation*}
  \end{small}
  \setlength{\abovedisplayskip}{3pt}
Thanks to $C10$, ${b_{i,k}}\frac{{{D_i}}}{{{R_{i,k}}}}$ is upper bounded by ${\chi _i}$ for all feasible solutions. By replacing each term ${b_{i,k}}\frac{{{D_i}}}{{{R_{i,k}}}}$ in the objective function of ${\cal P}2$ with its upper bound $b_{i,k}{\chi _i}$, we obtain the following convex approximation of ${\cal P}2$.\footnote{In the simulations, we observe that $C10$ is satisfied with equality with high probability, which implies that ${\cal P}2 - 1$ is a good approximation of ${\cal P}2$. }
\vspace{-1mm}
\begin{small}
\begin{equation*}\label{funcP2-1}
\begin{aligned}
{\cal P}2&-1:\mathop {min}\limits_{{\bf{W,P}}} {\rm{  }}\bm\zeta ({\bf{b}}{},{\bf{W}},{\bf{P}}) = \sum\limits_{i \in {\cal{I'}}} {\sum\limits_{k \in \cal{K}} {b_{i,k}} } {\chi _i}\sum\limits_{n \in \cal{N}} {{w_{i,n,k}}{p_{i,n,k}}} \\
s.t.& C4 - C6,{\rm{ }}C10{\rm{         }}.
\end{aligned}
\end{equation*}
\end{small}
{Obviously, ${\cal P}2 - 1$ is almost a strictly convex problem, except for the discrete subcarrier assignment value ${w_i}_{,n,k}$.  Relaxing ${w_i}_{,n,k}$ to be continuous between $\left[ {0,1} \right]$, the Lagrangian function is presented in (10) as shown at the top of the page, }
where ${\bm{\alpha }} = \left\{ {{\alpha _{i,k}},\forall i \in {{\cal I}'},\forall k \in {\cal K}} \right\}$ and ${\bm{\beta }} = \left\{ {{\beta _{i}},\forall i \in {{\cal I}'}}\right\}$ are Lagrange multiplier variables. It is worth noting that since ${\bf{W}}$ only assumes binary values and the constraint $C6$ implies that only a single user is assigned to each subcarrier. As a consequence, the constraints $C5$ is implicitly satisfied. Furthermore, the condition $\sum\limits_{k \in \cal {K}}\sum\limits_{n \in \cal{N}} {{w_{i,n,k}}{p_{i,n,k}}}  \ge 0$ is satisfied by the assumption that ${\bf{P}}$ is a positive variable matrix. Thus, the conditions $C5-C6$ can be omitted in the Lagrangian function (10).

For fixed ${{\bf{b}}}$, we may solve the problem ${\cal P}2 - 1$ in order to obtain the optimal power and subcarrier allocation strategies. Therefore, the following condition is both necessary and sufficient for the power allocation's optimality:
\begin{small}
\begin{equation}\label{equa11}
\setcounter{equation}{11}
\frac{{\partial \Re ({\bm{\alpha }},{\bm{\beta }},{\bf{W}},{\bf{P}})}}{{\partial {p_{i,n,k}}}}\left| {_{{p_{i,n,k}} = p_{i,n,k}^*}} \right. = 0,
\end{equation}
\end{small}
Then we can get the optimal power of user ${i}$ on subcarrier ${n}$ by (12),
\vspace{-3mm}
\begin{small}
\begin{equation}\label{equa13}
p_{i,n,k}^* ={\left[ {{\frac{{{\alpha _{i,k}}{B_N}}}{{\ln 2\left( {b_{i,k} {\chi _i} + {\beta _{i}}} \right)}} - \frac{{{\sigma ^2}}}{{{g_{i,n,k}}}}}} \right]^ + }.
\end{equation}
\end{small}
Once the optimal power allocation { ${\bf{P}^*(b)}$ }is calculated, the optimal subcarrier allocation strategy can be obtained through
\begin{small}
\begin{equation}\label{equa14}
\frac{{\partial \Re ({\bm{\alpha }},{\bm{\beta }},{\bf{W}},{\bf{P}^*(b)})}}{{\partial {w_{i,n,k}}}}\left| {_{{w_{i,n,k}} = w_{i,n,k}^*}} \right. = 0,
\end{equation}
\end{small}where the expression on the left side of the equation (13) is shown at the top of the page which denoted by (14).
\begin{small}
\begin{figure*}[ht]
\begin{align*} \label{lagrange}
\!\!\!\!\Re ({\bm{\alpha }},{\bm{\beta }},{\bf{W}},{\bf{P}}) \!= \!\sum\limits_{i \in {\cal{I'}}} {\sum\limits_{k \in \cal{K}} {b_{i,k}} } {\chi _i}\sum\limits_{n \in \cal {N}} {{w_{i,n,k}}{p_{i,n,k}}} \! + \!\sum\limits_{i \in {\cal{I'}}} {\sum\limits_{k \in \cal {K}} {{\alpha _{i,k}}} } \left( {b_{i,k}\frac{{{D_i}}}{{{\chi _i}}} \!-\! {R_{i,k}}} \right)\! +\! \sum\limits_{i \in {\cal{I'}}} { {{\beta _{i}}\left( \sum\limits_{k \in \cal{K}} {\sum\limits_{n \in \cal{N}} {{w_{i,n,k}}{p_{i,n,k}} \!-\! {P^m}} } \right)} }
\tag{10} 
\end{align*}
\hrulefill 
\end{figure*}
\end{small}Since the derivative in (14) is independent of $w_{i,n,k}$, it means that either the optimal value occurs at the boundaries of the feasible region or the derivative is null and hence the optimal subcarrier allocation is obtained inside the feasible region. Recalling $C6$ that each subcarrier is assigned to only one user, the optimal\ subcarrier allocation strategy ${\bf{W}^*(b)}$ is ruled by (15),
\begin{equation}\label{equa16}
\setcounter{equation}{15}
{w_{i,n,k}} = \left\{ \begin{array}{l}
  1,\quad\text{\rm{   if }}{\phi _{i,n,k}}\text{\rm{ =  min(}}{\bm{\phi} _{{\rm{ }}n}\bf(b))}\\
  0,\quad\text{\rm{   otherwise}}
\end{array} \right.,
\end{equation}
where ${\bm{\phi} _{{\rm{ }}n}\bf(b)}=\{\phi_{i,n,k}, i\in{\cal{I'}}\}, n\in {\cal{N}}$ is a row vector for lagrangian partial derivative on subcarrier ${n}$.
The dual variable matrices ${\bm{\alpha }}$ and ${\bm{\beta }}$ are updated using their corresponding subgradients,
\vspace{-1mm}
\begin{small}
\begin{equation}\label{equa17}
\begin{aligned}
{\alpha _{i.k}}(m + 1) = {\alpha _{i.k}}(m) + {\mu _\alpha }\left( {b_{i,k}\frac{{{D_i}}}{{{\chi _i}}} - {R_{i,k}}} \right),
\end{aligned}
\end{equation}
\end{small}
\vspace{-2mm}
\begin{small}
\begin{equation}\label{equa18}
\begin{aligned}
{\beta _{i}}(m + 1) =\! {\beta _{i}}(m) \!+\! {\upsilon _\beta }\left( \sum\limits_{k \in \cal{K}} {\sum\limits_{n \in \cal{N} } {{w_{i,n,k}}{p_{i,n,k}} \!-\! {P^m}} } \right),\!
\end{aligned}
\end{equation}
\end{small}where ${\mu _\alpha }$ and ${\upsilon _\beta }$ are the appropriate step sizes of the subgradient algorithm.
\vspace{-4mm}
\subsection{Upper-level Problem}
{Given the power and subcarrier allocation strategies $({\bf{W}^*(b)},{\bf{P}^*(b)})$ for fixed ${{\bf{b}}}$ as obtained from the algorithm in Section IV.A, the upper-level problem can be written as ${\cal P}3$,}
\vspace{-3mm}
\begin{small}
\begin{equation*}\label{funcP3}
\begin{aligned}
{\cal P}3&:\mathop {min}\limits_{{\bf{b}}} {\rm{  }}F({\bf{b}},{\bf{W}^*(b)},{\bf{P}^*(b)}) = \sum\limits_{i \in {\cal{I'}}} {\sum\limits_{k \in \cal {K}} {{b_{i,k}}E_{i,k}^r} } \\
s.t.& C1 - C3,{\rm{ }}C8.
\end{aligned}
\end{equation*}
\end{small}where the optimization problem is convex with respect to $b_{i,k}$, then the optimal offloading strategy ${{\bf{b}}^*}$ could be found once it satisfies the constraints $C1-C3$ , which can be represented as follows,
\vspace{-1mm}
\begin{small}
\begin{equation}\label{equa20}
{{\bf{b}}^*} \!= \!\arg \mathop {min}\limits_{{\bf{b}}} {\rm{  }}F({\bf{b}},{\bf{W}^*(b)},{\bf{P}^*(b)}) \!= \! \sum\limits_{i \in {\cal{I'}}} {\sum\limits_{k \in \cal{K}} {{b_{i,k}}E_{i,k}^r} }. \\
\end{equation}
\end{small}
The optimal offloading strategy in (18) becomes the minimal cost searching problem where the Hungarian method \cite{Refer16} can also be utilized to find a good solution. 

\vspace{-3mm}
\subsection{EEJS Algorithm}
Based on the above analysis of the bi-level problem, the EEJS algorithm is illustrated in \textbf{Algorithm 1}, which involves the solutions for the upper-level and lower-level problems.
{As seen from Algorithm 1, the lower-level problem possesses the computational complexity with proportional to the parameters $I'$, $K$ and $N$, while the upper-level problem has computational complexity in the order of ${\cal{O}}({KI'})$.

\begin{table}[t]
        \small
		\renewcommand{\arraystretch}{1.0}
		\label{table1}
		\begin{tabular}{p{75mm}}
			\hline
			\textbf{Algorithm 1: Energy-Efficient Joint Strategy (EEJS)}\\
			\hline
			\textbf{Input: ${\bf{b}}$, $\varepsilon $, ${I_{dd}}$, ${O_{I',K}}$}\\
            \textbf{Output: ${{\bf{P}}^*}$, ${{\bf{W}}^*}$, ${{\bf{b}}^*}$}\\
            \textbf{begin:}\\
            1.	\textbf{Initialize  ${\bf{P}}$, ${\bm{\phi}}$}\\
            2.	\textbf{for $o$ from} 1 \textbf{to} ${O_{I',K}}$\\
            3.	$m \leftarrow 0$;\\
            4.	\textbf{while} $m \le {I_{dd}}$ \textbf{or} $\left| {p_{i,n,k,o}^*(m + 1) - p_{i,n,k,o}^*(m)} \right| > \varepsilon$ \\
            5.	\textbf{for} $n$ \textbf{from} 1 to $N$\\
            6.	\textbf{for} $i$ \textbf{from} 1 to $I'$\\
            7.	\textbf{Compute $p_{i,n,k,o}^*$ according to (12);}\\
            8.	\textbf{Obtain the optimal subcarrier allocation $w_{i,n,k,o}^*$ using (14) and (15);}\\
            9.	\textbf{end for}\\
            10.	\textbf{end for}\\
            11. \textbf{Update the dual variables using (16) and (17);}\\
            12.	$m \leftarrow m + 1$;\\
            13.	\textbf{end while}\\
            14.	\textbf{end for }\\
            15.	\textbf{Find the optimal offloading strategy for (18);}\\
            \hline
            ${\bf{P}}$= initial power allocation matrix;\\
            ${\bm{\phi}}$= initial value matrix of (14);\\
            ${O_{I',K}}$= total offloading strategy types for $I'$ users and $K$ MEC servers;\\
            $p_{i,n,k,o}^*$= optimal power allocation at $o$th offloading strategy;\\
            $w_{i,n,k,o}^*$= optimal subcarrier allocation at $o$th offloading strategy;\\
            $\varepsilon$= power allocation precision;\\
            ${I_{dd}}$= maximum number of iterations;\\
            \hline
            \end{tabular}
\end{table}
\vspace{-3mm}

\section{Performance Evaluations}
\begin{spacing}{0.95}
In this section, we evaluate on the performance of the proposed algorithm and compare with conventional schemes.
Due to the fluctuation of channel state and restricted delay request, the task offloading strategy risks to fail in some situations. Accordingly, {successful offloading probability (SOP) is used to measure the reliability of the algorithm, which refers to the proportion of users who have successfully found appropriate MEC servers for task execution within the requested task delay threshold.}
\vspace{-3mm}
\subsection{System Setting}
 Mobile users and MEC servers are randomly deployed in a circle area which radius is 60 meters. The large scale fading of the channels is modeled as $PL = {d_{i,k}}^{ - \theta }$, where ${d_{i,k}}$ is the distance between user ${i}$ and MEC server ${k}$ and ${\theta}=2$ is the path-loss exponent. The Rayleigh fading model is adopted to model the small scale fading and $k_0$ is set to be $1 \times {10^{ - 24}}$ \cite{Refer13},\cite{Refer17}. In \cite{Refer18}, the energy consumption of the MEC server is ${10^{ - 5}}$ Joule/bit, and thus we set $k_1$ to be $1 \times {10^{ - 26}}$ to maintain the energy consumption per bit at the same order of magnitude. Besides, other simulation parameters employed in the simulations, unless otherwise mentioned, are summarized in TABLE I. 
\vspace{-3mm}

\subsection{Simulations}
In this part, performance of the proposed scheme is tested from aspects of the total energy consumption and SOP. Besides, two offloading baselines are presented for comparison.

\textbf{Baseline 1: minimum distance offloading algorithm (MDOA)}. This scheme indicates that the mobile users select the nearest MEC server to offload their tasks.

\textbf{Baseline 2: random offloading algorithm (ROA)}. It means each user randomly chooses a server to offload task.



	\begin{figure}[!t]\label{fig2}
		\centering
		\includegraphics[width=7.2cm,height=5.0cm]{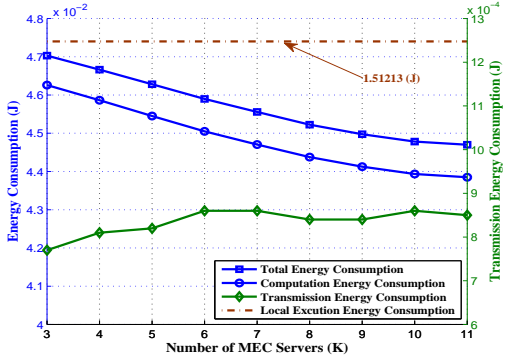}
		\vspace{-0.15em}
		\caption{Energy Consumption w. r. t. Total Number of MEC Servers. $I'$ = 3, $N$ = 64.}
	\end{figure}

\subsubsection{Performance of the Proposed EEJS Algorithm}
  The total energy consumption and corresponding computation/transmission energy consumption of our proposed algorithm are shown in Fig. 2. As the number of MEC servers increases, it is observed that the computation energy consumption takes similar trend with the total energy consumption that they decrease continuously towards stability. Conversely, the energy consumption for data transmission is a growing trend. These phenomena can be mutually explained as 1) the energy consumption of computation takes a large majority part of the total energy consumption; 2) in order to minimize the total energy consumption, the optimal offloading strategy tends to exploit multi-MEC diversity while sacrificing multi-user diversity to some extent.
 Meanwhile, the dashed line indicates the energy consumption for the local computation execution case, and we can see that our proposed edge offloading strategy saves considerable energy especially when there are plenty of surrounding MEC servers.
\begin{table}[tbp]
\small
\caption{\textbf{Simulation Parameters}}  \centering \vspace{-0.2cm}
\renewcommand{\arraystretch}{0.8}
\label{table2}
\begin{tabular}{p{4.7cm}p{3.0cm}ll}
\hline
\textbf {MEC System Parameters} & \textbf{Values} \\
\hline
\noalign{\smallskip}
Subcarrier bandwidth $B_N$ &12.5 kHz \\
Background noise ${\sigma ^2}$ &-113 dBm \\
Maximum transmit power $P^m$  &600 mW \\
Input data size ${D_i},i\in{\cal{I'}}$  &1000-1100 bits \\
The computation workload/intensity ${X_i},i\in{\cal{I'}}$  &1000-1200 (cycles/bit) \\
Task deadline ${\tau _i},i\in{\cal{I'}}$  &9-10 ms \\
The CPU frequency of mobile users ${f_{i,loc}},i\in{\cal{I'}}$  &0.6-0.7 GHz \\
The CPU frequency of MEC servers ${f_{k,ser}},k\in{\cal{K}}$  &1.1-1.2 GHz \\
\hline
\textbf {Lagrange Iteration Parameters} & \textbf {Values} \\
\hline
\noalign{\smallskip}
Maximum number of iterations ${I_{dd}}$ &600 \\
Power allocation precision $\varepsilon $  &${10^{ - 5}}$ \\
Step sizes ${\mu _\alpha }$ &$2 \times {10^{ - 18}}$ \\
Step sizes ${\upsilon _\beta }$  &${10^{ - 5}}$ \\
\hline
\end{tabular}
\end{table}


	\begin{figure}[!t]\label{fig3}
		\centering
		\includegraphics[width=7.2cm,height=5.0cm]{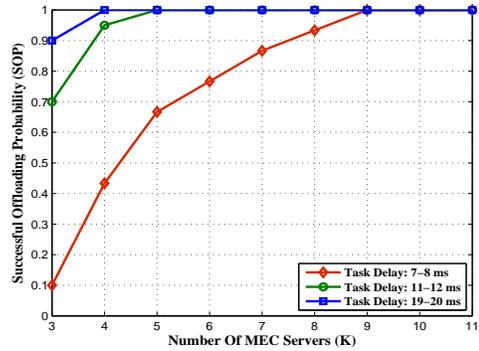}
		\vspace{-1mm}
		\caption{Successful Offloading Probability (SOP) w. r. t. Total Number of MEC Servers. $I'$ = 3, $N$ = 64.}
	\end{figure}

 The SOP for different task delay constraints with the increase of the MEC server number is shown in Fig. 3. For the delay-sensitive tasks, the SOP is lower.
  The more MEC servers distributed in the MEC system or the looser deadline restriction for tasks, the larger successful probability users could find the optimal offloading target server. Obviously, more MEC servers provides more offloading targets for mobile users and the SOP is higher naturally. The deadline restriction is the dominant constraint and influencing factor of SOP, the looser deadline restriction implies the looser requirement for a series of influence factors including channel state, transmission power and computation resources and thus results in higher SOP.
\begin{figure}[!t]\label{fig5}
	\centering
	\includegraphics[width=7.2cm,height=5.0cm]{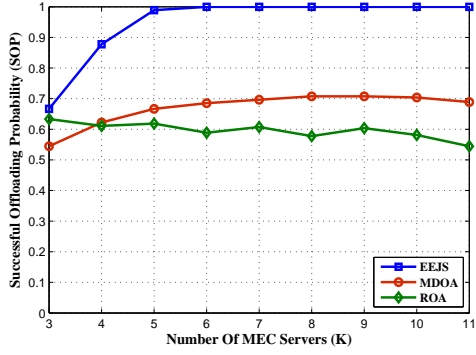}
		\vspace{-0.15cm}
	\caption{Successful Offloading Probability (SOP) w. r. t. Total Number of MEC Servers. $I'$ = 3, $N$ = 64. Legends: EEJS - energy-efficient joint strategy, ROA - random offloading algorithm, MDOA - minimum distance offloading algorithm.}
\end{figure}

\vspace{-0.4em}
\begin{figure}[!t]\label{fig4}
	\centering
	\includegraphics[width=7.2cm,height=5.0cm]{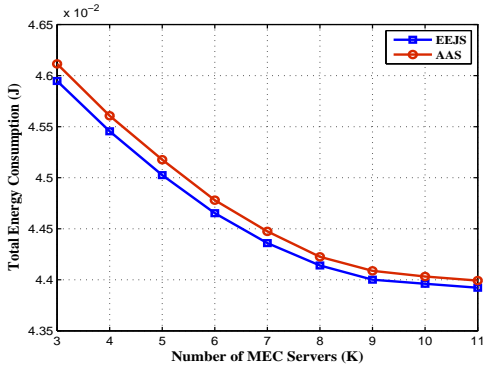}
	\vspace{-0.15cm}
	\caption{Total Energy Consumption w. r. t. Total Number of MEC Servers. $I'$ = 3, $N$ = 60. Legends: EEJS - energy-efficient joint strategy, AAS - average allocation strategy.}

\end{figure}
\vspace{1mm}
\subsubsection{Comparison with the Referencing Offloading Algorithms}

The SOP for different algorithms is shown in Fig. 4, from which we observe that our proposed algorithm obtains the highest SOP compared with the other algorithms especially for the larger number of MEC servers, which implies that by combining the wireless resource and computing offloading, the EEJS achieves the tradeoff between multi-user diversity and multi-MEC diversity. Meanwhile, we can easily find out MDOA is better than ROA as the number of MEC servers increases. This is because that MDOA selects the nearest MEC servers as the offloading anchors, thus it gains in the multi-user diversity than ROA.

In addition, the average allocation strategy (AAS), which allocates power and subcarrier equally is compared for the energy consumption with our proposed EEJS algorithm. As shown in Fig. 5, our proposed algorithm advances in the energy consumption especially in the sparse network. With the number of MEC servers increasing, the computing resources in the network become more abundant, so mobile users will sacrifice transmission energy consumption to search for better computing resources. Consequently, the wireless resource allocation has weak effect on the energy saving in this situation, and thus we can find the gap between the energy consumption of these two algorithms decreases when the number of MEC servers is large.

\vspace{-5mm}

\section{Conclusions}
In this work, we discussed the joint optimization of offloading strategy and wireless resource allocation problem in a MEC system with multiple users and servers. In the resource constrained system, an EEJS algorithm exploiting the tradeoff between the the multi-user diversity and multi-MEC diversity is proposed. Simulation results reveal that our proposed algorithm achieves the better performance in energy consumption and SOP.
\end{spacing}

%
%
%

\end{document}